\documentclass[sigconf,natbib=true]{acmart}
\AtBeginDocument{%
  }

\usepackage{xcolor}   
\usepackage[normalem]{ulem}  
\usepackage{soul}     

\newcommand{\ours}{STAR}
\usepackage{enumitem}

\usepackage{amsfonts}
\usepackage{booktabs}
\usepackage{multirow}
\usepackage{makecell}
\usepackage{tabularx}
\usepackage{subcaption}
\usepackage{array}
\usepackage{siunitx}
\usepackage{algorithm}
\usepackage[noend]{algorithmic}

\begin{document}

\title{STAR: Semantic-Tuned and Tail-Adaptive Retriever for Graph-Augmented Generation}

\author{Shuai Li, Chen Huang, Duanyu Feng, Wenqiang Lei, See-Kiong Ng}



\begin{abstract}
To augment Large Language Models (LLMs) for multi-hop question answering, a mainstream solution within Graph Retrieval Augmented Generation (GraphRAG) leverages lightweight retrievers to efficiently extract information from a given Knowledge Graph (KG). However, existing methods often overlook the inherent challenge of sparse semantic information in graphs. Specifically, our experiments reveal that these methods produce biased retrieval Semantic Shortcut Bias and Long-Tail Path Bias, leading to inadequate semantic modeling and limited GraphRAG effectiveness. To address these issues, we propose \ours, a semantic-tuned and tail-adaptive retriever for GraphRAG. \ours~integrates two key learning paradigms: token-level interaction learning and path-weighted contrastive learning. The former employs a cross-attention architecture and a hard path mining mechanism to jointly model the query and path, thereby mitigating the Semantic Shortcut Bias. The latter introduces a tailored contrastive learning objective that utilizes tail-adaptive path weighting, designed to optimize the training process and ease the Long-Tail Path Bias. Extensive experiments demonstrate that \ours~consistently outperforms baselines, achieving average retrieval performance gains of 1.8\% and LLM QA performance improvements of 2.2\% across all benchmark datasets. Our code is available at \url{https://anonymous.4open.science/r/STAR-C583}.
\end{abstract}





\maketitle



\section{Introduction}


Traversing a Knowledge Graph (KG) to retrieve nodes has emerged as a prominent method (i.e., GraphRAG) to enhance Large Language Model (LLM) performance in complex, multi-hop reasoning tasks \cite{hu-etal-2025-grag, peng2024graph}, thereby mitigating hallucinations in web applications \cite{sen-etal-2023-knowledge}. While prompting a LLM for graph traversal presents an alternative, it is generally associated with substantially increased invocation expenditures and diminished retrieval efficacy \cite{jin-etal-2024-graph, sun2024thinkongraph, luo2024reasoning}. In response, employing lightweight modules for retrieving the relevant path on the graph (we term it as \textit{Lightweight Retriever}) is gaining increasing attention \cite{RD-P, zhang2024knowgptknowledgegraphbased}, marking a seminal shift within GraphRAG research.

\begin{figure}
  \centering
    \includegraphics[width=0.38\textwidth]{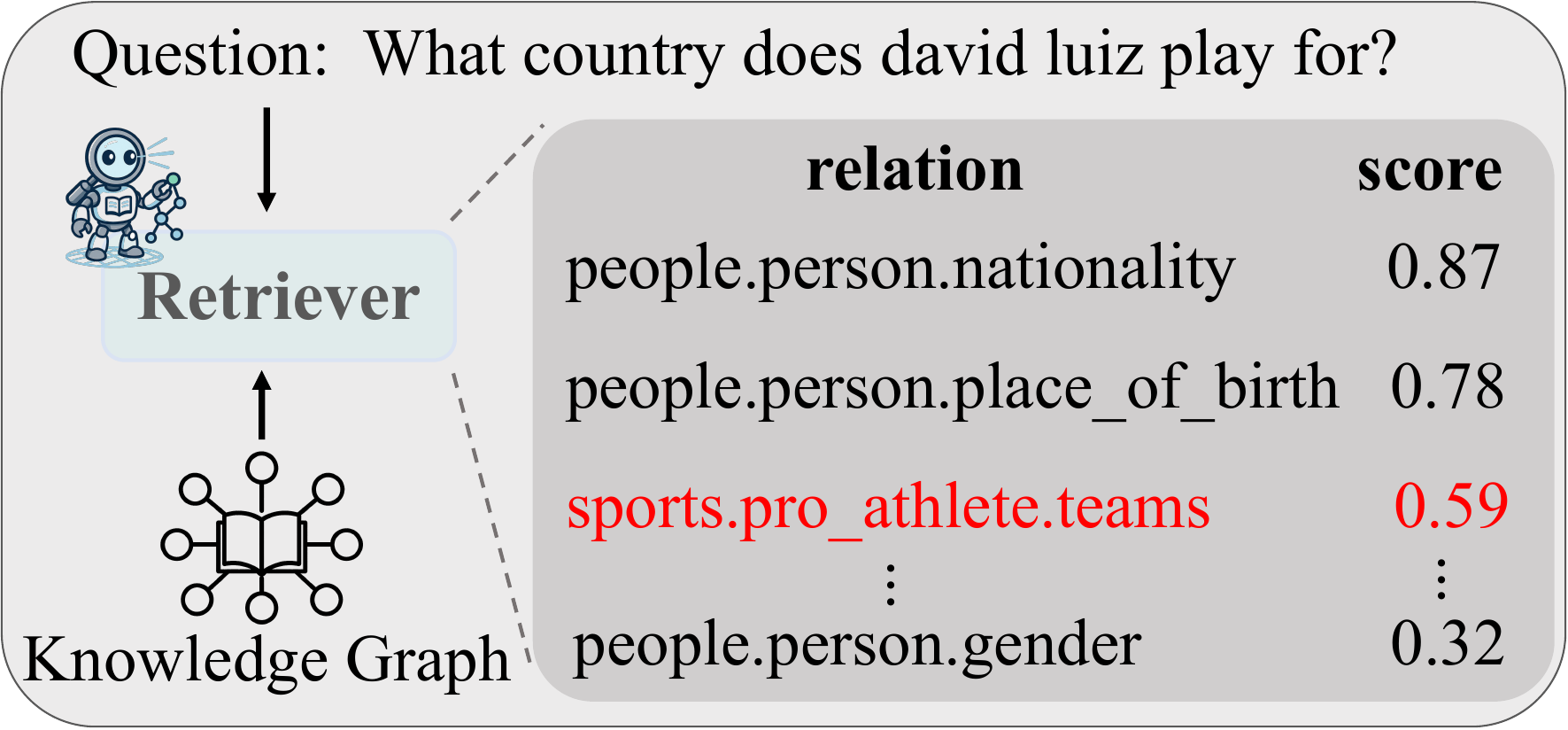}
    \setlength{\abovecaptionskip}{0pt}
\setlength{\belowcaptionskip}{0pt}
  \caption{Illustrations of semantic shortcut, where a retriever prefers a semantically similar but incorrect relation.}
  \label{fig:overall-bias-case}
  \vspace{-3mm}
\end{figure}


Unlike document-oriented retrieval's rich textual semantics for user queries, lightweight GraphRAG retrievers face the inherent challenge of sparse semantic information in graph. These semantics are often limited to node names (e.g., \textit{London}) or sequential path information (e.g., \textit{\{London, place\_of\_birth, Charlie Chaplin, place\_of\_burial, ...\}}). Consequently, a central research objective for lightweight GraphRAG retrievers is to effectively encode richer semantic information from the graph, ensuring accurate traversal for LLM enhancement. To this end, lightweight retriever development has progressed through several stages. Specifically, early approaches applied traditional graph algorithms, such as Breadth-First Search (BFS), for graph traversal \cite{li-ji-2022-semantic, NEURIPS2024_efaf1c97}, while subsequent studies introduced neural networks like Multi-Layer Perceptrons (MLPs) as scoring functions to model query–path semantic similarity on KGs \cite{hu-etal-2025-grag}. While efficient, these methods offer only a shallow semantic understanding. Consequently, the current dominant paradigm employs Pre-trained Language Models (PLMs) to capture the nuanced semantics embedded within the graph \cite{baek-etal-2023-knowledge, munikoti2023atlanticstructureawareretrievalaugmentedlanguage, RD-P}, utilizing PLMs as dual encoders, independently encoding the user query and each graph path to optimize their similarity.

\begin{figure}[t]
  \centering
    \includegraphics[width=0.4\textwidth]{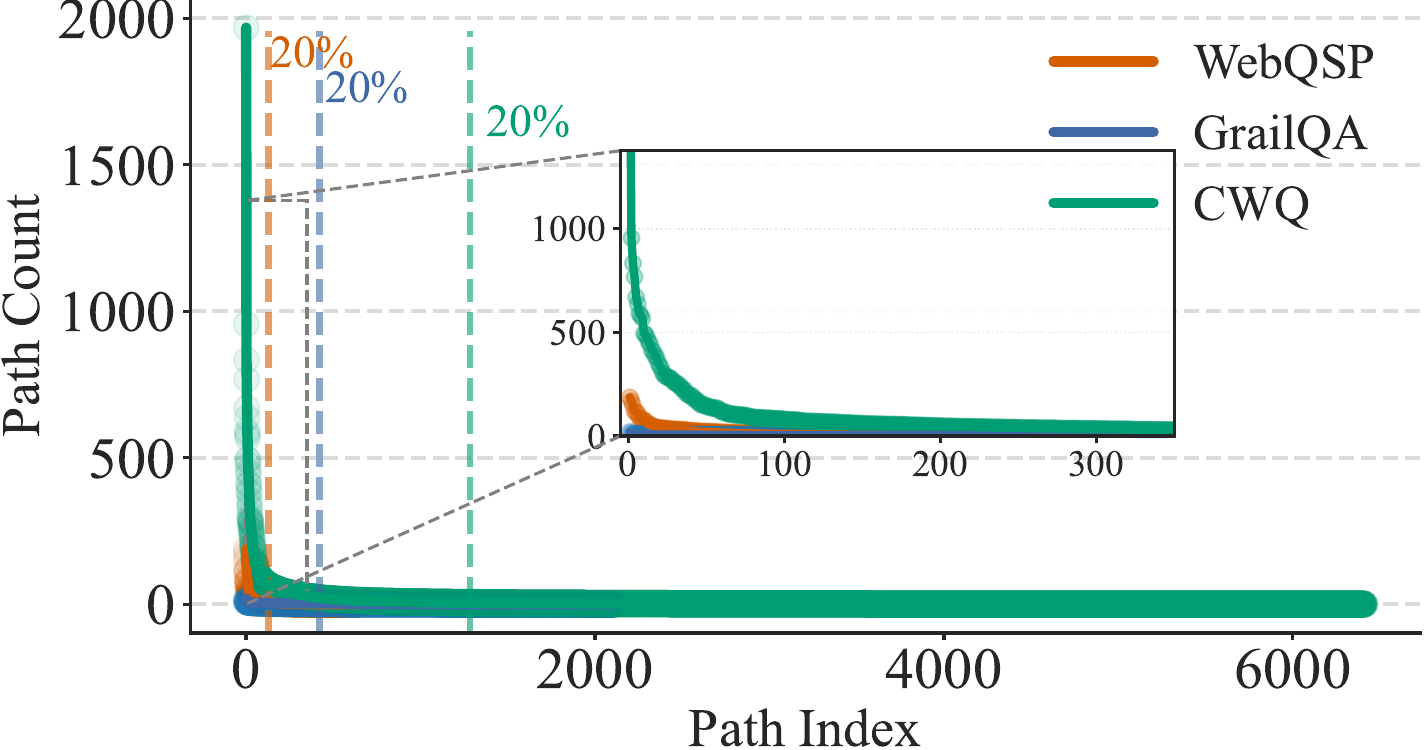}
  \caption{Long-tail distribution of path frequencies.}
  \label{fig:overall-bias-case2}
  \vspace{-3mm}
\end{figure}

Despite showing performance improvements, these existing methods still overlook the sparse semantic nature of KGs, leading to unsatisfactory semantic modeling and, consequently, limited system effectiveness. In particular, 1) \textbf{Semantic Shortcut Bias}. The inherent sparse semantic nature of KGs demands a more fine-grained understanding between user queries and graph paths from lightweight retrievers, enabling better utilization of the graph's available information. Existing methods, by independently encoding queries and paths as isolated wholes to then force similarity learning. This prevents the crucial token-level interactions required for fine-grained logical understanding. 
Taking Figure~\ref{fig:overall-bias-case} for example, given the query `\textit{What country does David Luiz play for?}', since `\textit{play for}' implies club membership, it necessitates to search for the team first (i.e., the relation \textit{sports.pro\_athlete.teams}) before retrieving the country. However, current methods often rely on superficial semantic similarities. They tend to match  `\textit{David Luiz}' with \textit{people.person} and `\textit{country}' with \textit{nationality}, directly selecting the logically incorrect relation \textit{people.person.nationality}. We refer to this error as \textit{Semantic Shortcut Bias}.
This highlights the need for models to perform token-by-token comprehension between query and path to discern subtle contextual differences.
2) \textbf{Long-Tail Path Bias}. The practical usage of semantic information within KGs often exhibits a sparse, highly skewed distribution in user queries. This means a few path types dominate in frequency within many training datasets, while the vast majority appear rarely (Figure~\ref{fig:overall-bias-case2}). This frequency-based semantic sparsity introduces an additional challenge for retriever learning. We experimentally show that ignoring this issue causes retrievers to become experts on these common `head' paths but consistently fail on the far more numerous `tail' paths (cf. Section \ref{preli}). This implies compromised robustness and reduced performance on queries that necessitate less frequent paths.

To address these biases,we propose the \underline{S}emantic-tuned and \underline{T}ail-\underline{A}daptive \underline{R}etriever for GraphRAG (\textbf{\ours}), featuring token-level interaction learning and path-weighted contrastive learning to ease the two biases, respectively. Specifically, instead of separately encoded representations, retriever should jointly model the query and path to capture their mutual information, which manifests as token-level interactions. To achieve this, \ours~moves beyond separate query-path modeling by employing a single-tower PLM-based architecture. This architecture employs cross-attention to facilitate token-level interactions, thereby capturing fine-grained relational semantics between queries and candidate paths. Then, we follow the idea of hard data mining \cite{shrivastava2016training, smirnov2018hard, huang-etal-2024-selective-annotation} and enhance \ours~to learn to distinguish between correct paths and semantically similar yet incorrect distractors (i.e., hard-to-distinguish paths).
Subsequently, \ours~optimizes this architecture through path-weighted contrastive learning, employing a customized graph context to weight each training path, thereby emphasizing rare, long-tail paths. To further enhance this. As such, \ours~offers a unified retrieval solution that captures fine-grained query-path interactions to ease the semantic shortcut, while its path-weighted learning paradigm enhances robustness to skewed long-tail distributions in web-sourced KGs.

To evaluate the effectiveness of \ours, we conduct extensive experiments across various benchmark datasets, all derived from web-sourced knowledge graphs that simulate real-world Web interactions. Results demonstrate that \ours~ consistently outperforms baselines, achieving average retrieval performance gains of 1.8\% and LLM QA performance improvements of 2.2\% across all evaluated datasets. Further ablation analysis reveals that \ours's success is attributable to a \textbf{50.0\%} reduction in errors related to the semantic shortcut bias and a \textbf{14.1} reduction in those regarding the long-tail path bias. Additionally, we find that \ours~is robust across different backbones and LLMs, and adjusting the beam search width further enhances its performance, offering flexibility for diverse applications.
In summary, our main contributions are as follows:
\begin{itemize}[leftmargin=*]
    \item We identify and diagnose two critical biases in modern PLM-based lightweight retriever for GraphRAG, leading to unsatisfactory semantic modeling and GraphRAG effectiveness.
    
    \item We propose \ours, featuring token-level interaction learning and path-weighted contrastive learning to ease the two biases.
    
    \item We conduct extensive experiments showing \ours~achieves state-of-the-art results on web-sourced benchmarks, with analysis confirming it effectively mitigates both identified biases. 
\end{itemize}



\section{Related Works}
Our research is closely tied to the development of lightweight retrievers for GraphRAG, aiming to achieve efficient and semantically effective retrieval to augment LLMs. This section provides a literature review and highlight our differences.

\textbf{LLM-Centric Retrievers for GraphRAG}. 
Prompting the LLM to make graph traversal decisions and achieve iteratively next-hop selection, as a distinct type of GraphRAG method, have attracted considerable attention \cite{sun2024thinkongraph, dehghan-etal-2024-ewek}. Usually, this approach is often enhanced through advanced techniques such as Monte Carlo Tree Search \cite{tran2025rareretrievalaugmentedreasoningenhancement, huang2024ritekdatasetlargelanguage, qi2025mutual} or ReAct \cite{yao2023react, li-etal-2024-graphreader, jin-etal-2024-graph}. 
However, they are frequently criticized for being prohibitively slow and cost-intensive for real-time applications, particularly when dealing with web-scale KGs. Furthermore, some necessitate extensive and costly fine-tuning to adapt their general abilities to the structured nature of knowledge graphs \cite{luo2024reasoning}, posing challenges for deployment and maintenance in dynamic Web environments.

\textbf{Lightweight Retrievers  for GraphRAG}. As an efficient solution, lightweight retrievers have garnered considerable favor.
Early work used traditional graph algorithms like BFS~\cite{li-ji-2022-semantic, NEURIPS2024_efaf1c97}, but these methods often lack semantic flexibility. Transitioning towards more semantically aware methods, some proposed using Multi-Layer Perceptrons (MLPs) to score query-path similarity for efficient pruning and path selection~\cite{hu-etal-2025-grag}. This progression ultimately led to the current standard: employing Pre-trained Language Models (PLMs) for high-speed retrieval.
The most common implementation is the embedding similarity based retrieval, where query and path are encoded independently for fast similarity matching~\cite{baek-etal-2023-knowledge, munikoti2023atlanticstructureawareretrievalaugmentedlanguage, RD-P}. Some methods also apply reinforcement learning, using similarity as a reward signal to guide path selection~\cite{zhang2024knowgptknowledgegraphbased, shen2025insight}. 
However, existing methods often overlook the sparse semantic nature of KGs, leading to unsatisfactory semantic modeling and GraphRAG performance. In this regard, we experimentally reveal two critical biases and propose \ours~to achieve more effective performance.



\section{Preliminary Experiment}
\label{preli}

This section details the limitations of existing lightweight retrievers. Driven by the sparse semantic nature of KGs, our experimental analysis investigates two key aspects: first, the quantification of query-path similarity to analyze graph structure-based semantic sparsity; and second, the analysis of path frequency within training datasets to study frequency-based semantic sparsity. This investigation ultimately identifies two critical biases: the Semantic Shortcut Bias and the Long-Tail Path Bias. 

\subsection{Experimental Setup}
\label{psetup}
\noindent\textbf{Evaluation Overview.} We partition the retrieval results into biased and unbiased subsets to compare the prevalence of bias-related features across both groups. To analyze the Semantic Shortcut Bias, we compute the cosine similarity between retrieved paths and user queries using SimCSE\footnote{According to the original paper, SimCSE achieves a high correlation (83.76\%) with human judgment, making it a robust proxy for detecting semantic shortcuts.}~\cite{gao2021simcse} embeddings. Subsequently, we compare the percentage of paths with significant semantic similarity ($>0.95$) found in the correct retrieved paths and the incorrect ones.
For the Long-Tail Path Bias, we follow the Pareto principle~\cite{zhu2015pareto} by classifying the least frequent 20\% of paths in the training set as `\textit{tail paths}'. We then analyze the prevalence of these paths within both the correct and incorrect groups.
Finally, a higher prevalence of these features in the incorrect subset as evidence that the corresponding bias contributes to retrieval errors.

\noindent\textbf{Datasets}. Following previous studies \cite{RD-P, sun2024thinkongraph, hu-etal-2025-grag}, we consider three multi-hop QA benchmark datasets, including \texttt{WebQSP} \cite{WebQSP}, \texttt{CWQ} \cite{CWQ}, and \texttt{GrailQA} \cite{GrailQA}. These datasets collectively cover a wide spectrum of challenges, including multi-hop and compositional reasoning, as well as generalization across multiple distinct domains.

\noindent\textbf{Metrics}. Inspired by previous studies \cite{RD-P, sun2024thinkongraph, hu-etal-2025-grag}, we report the ratio of bias-affected samples in correct vs. incorrect retrievals.

\noindent\textbf{Baselines}. Following common practice in the field \cite{zhang2025survey}, we consider lightweight retrievers, \textit{G-Retriever} \cite{NEURIPS2024_efaf1c97}, \textit{GRAG} \cite{hu-etal-2025-grag} and \textit{RD-P} \cite{RD-P}, and even a LLM-centric retriever, \textit{RoG} \cite{jin-etal-2024-graph}. 

\subsection{Experimental Findings: Two Biases}
\label{sec:two-biases}
\begin{figure}[t]
  \centering
  \begin{subfigure}[t]{\linewidth}
    \centering
    \begin{minipage}{\linewidth}
      \centering
      \includegraphics[width=\linewidth]{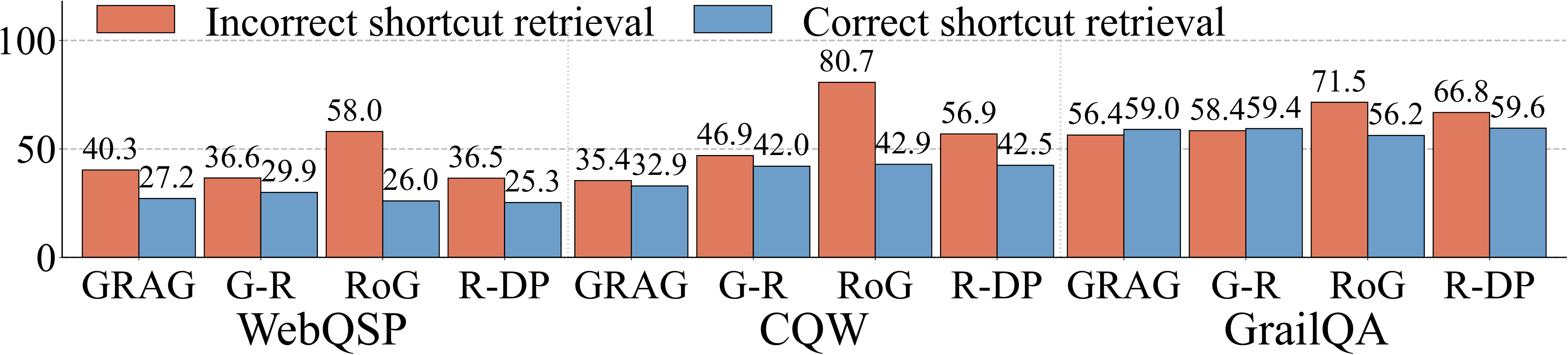}
    \end{minipage}
    \hfill
    \begin{minipage}{\linewidth}
      \centering
      \includegraphics[width=\linewidth]{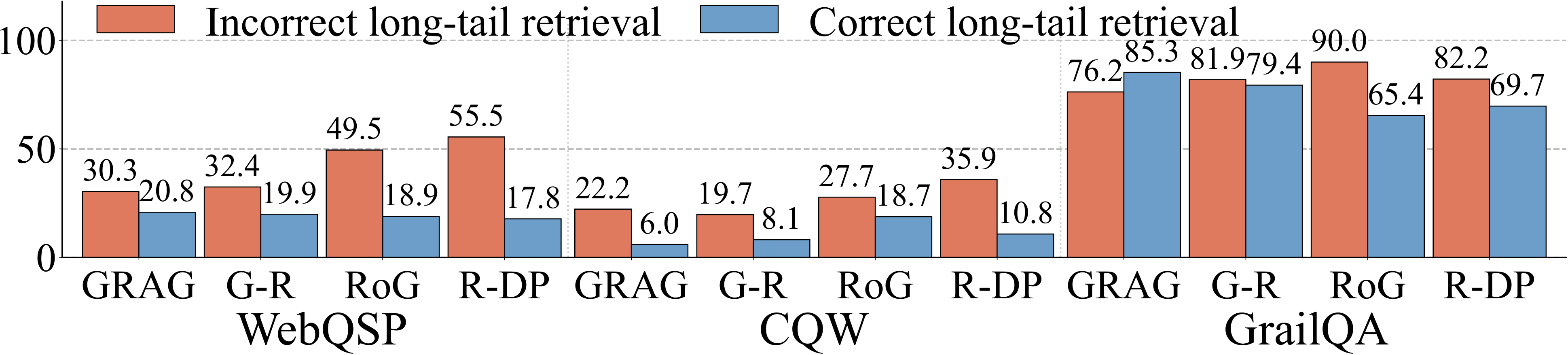}
    \end{minipage}
  \end{subfigure}

  \caption{Analysis on Semantic Shortcut (Top) and Long-Tail Path Bias (Bottom) on correct and incorrect retrieval subsets. X-axis represents the ratio value (\%).
  }
  
  \label{fig:overall-bias-statistics}
\end{figure}

Figure \ref{fig:overall-bias-statistics} presents our experimental results, the consistent disparity between incorrect and correct subsets confirms that both biases significantly contribute to retrieval failures.

\noindent\textbf{Semantic Shortcut Bias}. As shown in Figure~\ref{fig:overall-bias-statistics} (top),  
existing methods exhibit a biased path selection, where superficial textual similarity to the query is prioritized, resulting in erroneous retrievals. Specifically, on \texttt{WebQSP}, while \textit{RoG} only selects shortcuts in 26.0\% of correct predictions, this ratio surges to \textit{58.0\%} in wrong predictions. This \textit{32.0\% gap} clearly indicates that superficial high similarity acts as a distractor leading to errors. Similarly, on \texttt{GrailQA}, \textit{RD-P} shows a higher shortcut ratio in failures (66.8\%) than in successes (59.6\%). 
This validates that current methods struggle to distinguish fine-grained semantic differences when textual overlap is high.
This highlights that current retrieval methods, lacking fine-grained interaction, tend to prioritize paths that ``look'' similar to the query (lexical overlap) instead of those that are logically correct.
The prevalence of this bias underscores the significance of fine-grained query-path interactions.

\noindent\textbf{Long-Tail Path Bias.} 
The inherently skewed, long-tail distribution of paths in real-world KG training datasets (cf. Figure~\ref{fig:overall-bias-case2}) impedes the semantic learning of existing methods. 
As shown in Figure~\ref{fig:overall-bias-statistics} (bottom), the proportion of long-tail paths is markedly higher in the Wrong set for most baselines.
On \texttt{WebQSP}, \textit{RD-P} exhibits a massive disparity: \textbf{55.5\%} of its errors involve long-tail paths, compared to only \textbf{17.9\%} of its successes. This implies the model is well-fitted to head paths but fails disproportionately on tail paths.
On \texttt{GrailQA}, this trend is even more severe for \textit{RoG}, where long-tail paths account for \textbf{90.0\%} of errors versus 65.4\% of correct cases.
These results also highlight the significant challenge posed by Long-Tail Path Bias, even for powerful LLMs fine-tuned on KG reasoning tasks. This underscores the importance of managing paths with diverse distributions to ensure robust retrieval across varied queries.



\section{\ours}

\begin{figure*}[t] 
  \centering
  \includegraphics[width=0.98\textwidth]{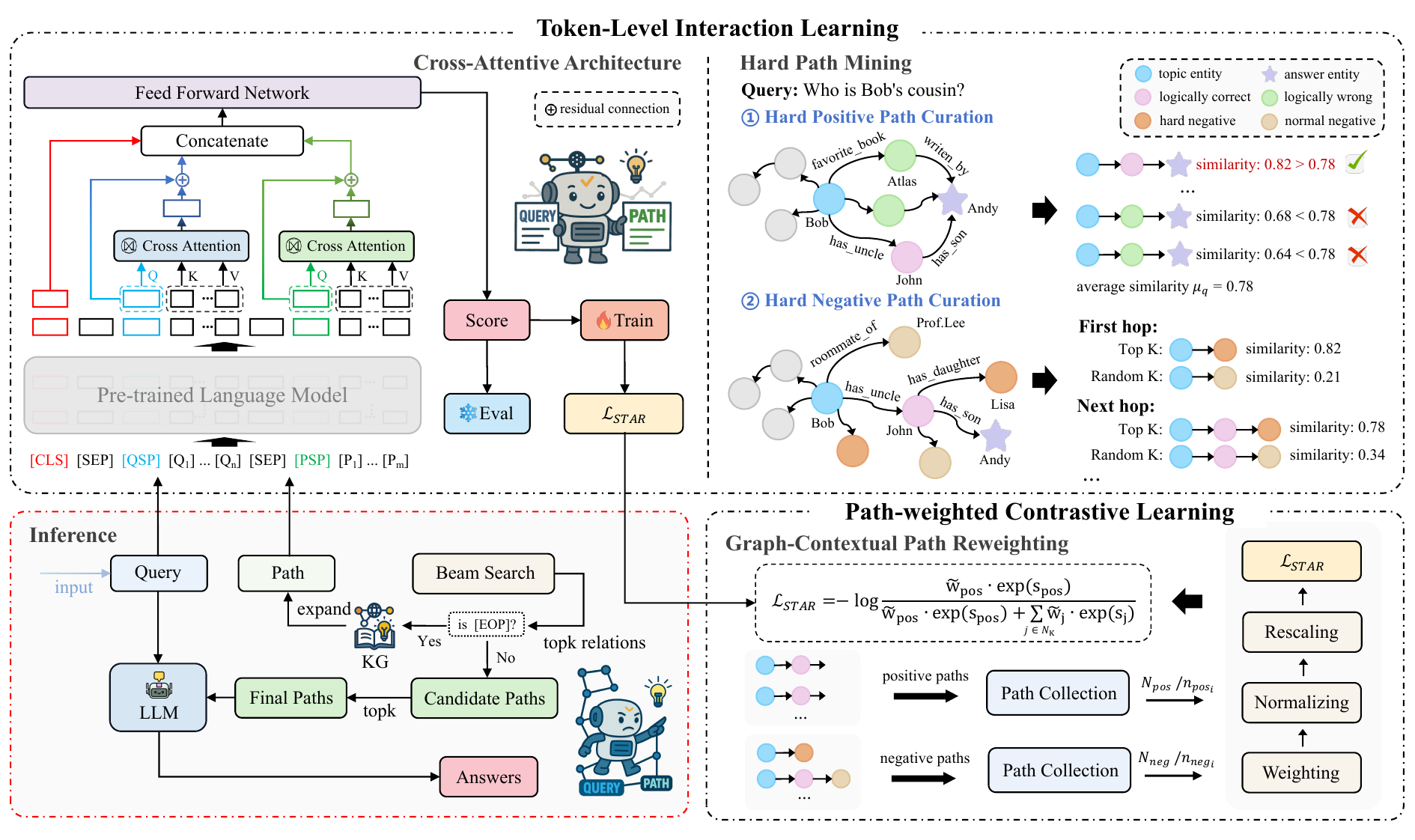}
  \setlength{\abovecaptionskip}{0pt}
\setlength{\belowcaptionskip}{0pt}
  \caption{Overview of our \ours, featuring token-level interaction learning and path-weighted contrastive learning to ease the two biases, respectively. Our cross-attentive architecture jointly models query-path pairs to capture fine-grained semantic interactions, optimized with a contrastive loss on hard-mined path samples. Simultaneously, our path-weighting scheme amplifies the learning signal from infrequent paths within the final loss $\mathcal{L}_{\text{\ours}}$, to ensure robust performance across the entire data distribution. During inference, the trained STAR model guides a beam search to retrieve paths for final answer generation.}
  \label{fig:framework}
\end{figure*}

\noindent\textbf{Overview of \ours}. In response to the two biases, \ours, as shown in Figure~\ref{fig:framework}, incorporates two learning paradigms: token-level interaction learning and path-weighted contrastive learning. The former utilizes a cross-attention architecture and a hard path mining mechanism to jointly model the query and path, thereby ensuring fine-grained interactions. The latter introduces a tailored contrastive learning that employs tail-adaptive path weighting, designed to optimize the training process of \ours. Each of these components is detailed in the subsequent section (Section \ref{part1} and Section \ref{sec:LOSS}).

\noindent\textbf{Task Definition}. In this paper, our task is formulated within the standard GraphRAG framework. A knowledge graph is formally defined as a set of triples $\mathcal{G} = \{(e, r, e') \mid e, e' \in \mathcal{E},\ r \in \mathcal{R} \}$, where $\mathcal{E}$ and $\mathcal{R}$ represent the sets of entities and relations, respectively. Given a knowledge graph, a path $p$ is denoted as a sequence of alternating entities and relations on the graph: $p = e_t \xrightarrow{r_1} e_1 \dots \xrightarrow{r_l} e_l$. This path originates from a topic entity $e_t$ derived from the query $q$ using extraction techniques such as LLMs or Named Entity Recognition (NER), and terminates at an answer entity $e_l$. Consequently, for a given query $q$, the primary objective of a lightweight retriever is to identify the optimal path $p^*$. This retrieval objective entails learning a model $p_\theta$ that assigns scores to candidate paths as follows:
\[
p^* = \arg\max_{p \in \mathcal{P}} p_\theta(p \mid q, \mathcal{G}),
\]
where $\mathcal{P}$ signifies the set of all candidate paths. The identified optimal path $p^*$ is subsequently fed into a LLM to produce the final answer. Our research primarily focuses on enhancing the retriever $p_\theta$ to ensure the delivery of high-quality knowledge augmentation.


\subsection{Token-Level Interaction Learning}
\label{part1}
Unlike previous PLM-based approaches \cite{RD-P, shen2025insight} that encode query and path separately, \ours~is designed for fine-grained interaction between the query and candidate paths. To achieve this, \ours~is instantiated as a single-tower cross-attentive architecture that scores paths through fine-grained interaction with query. 

\textbf{Input of \ours}. 
Given a PLM backbone\footnote{The effectiveness of \ours~is robust across various PLM backbones; see Section \ref{ana1}.}, such as RoBERTa \cite{gao2021simcse}, this PLM processes a joint representation of the query $q$ and a candidate path $p$. Note that this joint processing enables the model to capture their fine-grained relationships later on. Specifically, we construct the input sequence for the PLM, denoted as $\mathbf{x}$, by concatenating the $q$ and $p$, separated by special tokens, as follows:
\[
\mathbf{x} = \{[\text{CLS}], [\text{SEP}], [\text{QSP}], q, [\text{SEP}], [\text{PSP}], p\}.
\]
Here, $[\text{QSP}]$ and $[\text{PSP}]$ are special markers for the query and path segments, respectively. $\text{[CLS]}$ serves as the classification token for global context, and $\text{[SEP]}$ is the segment delimiter. 
The candidate path $p $ is constructed by concatenating the topic entity $e_t$ (identified from the query) and the relation path $r_1, r_2, \dots, r_l $, forming as $\{e_t, [\text{SEP}], r_1, \dots, [\text{SEP}], r_l\}$.

\textbf{Cross-attentive Architecture}. To achieve token-level interaction, \ours~employs an explicit cross-attention mechanism within the contextualized hidden state space of the PLM output. The design enables the $[\text{QSP}]$ and $[\text{PSP}]$ tokens to aggregate relevant information from their respective query and path sequences.
Specifically, let $h_{\text{[QSP]}} \in \mathbb{R}^d$ and $h_{\text{[PSP]}} \in \mathbb{R}^d$ be the hidden states of these special tokens, and $H_Q \in \mathbb{R}^{n \times d}$ and $H_P \in \mathbb{R}^{m \times d}$ denote the hidden states of the query and path tokens, respectively.
We then apply the following cross-attention operation:
\[
\text{Attn}(q, K, V) = \text{softmax}\left(\frac{W_Q(q) \cdot W_K(K)^T}{\sqrt{d_k}}\right) V,
\]
where $W_Q(\cdot)$ and $W_K(\cdot)$ represent two linear transformations, parameterized by learnable weight matrices $W_Q, W_K \in \mathbb{R}^{d \times d_k}$. Here, $d_k$ is the dimensionality of the transformed keys. In order to aggregate information for the query and the path, \ours~applies two distinct cross-attention operations to the $[\text{QSP}]$ and $[\text{PSP}]$ tokens, respectively, as follows:
\[
h'_{\text{[QSP]}} = \text{Attn}(h_{\text{[QSP]}}, H_Q, H_Q),
\quad 
h'_{\text{[PSP]}} = \text{Attn}(h_{\text{[PSP]}}, H_P, H_P).
\]

\noindent Residual connections is then applied to preserve original semantic information and enhance stability:
\[
h''_{\text{[QSP]}} = h_{\text{[QSP]}} + h'_{\text{[QSP]}}, \quad h''_{\text{[PSP]}} = h_{\text{[PSP]}} + h'_{\text{[PSP]}}.
\]
As such, these refined representations, $h''_{\text{[QSP]}}$ and $h''_{\text{[PSP]}}$, capture the essence of the query and path through their mutual token-level interaction. Finally, \ours~produces a single relevance score $s$ by concatenating these two representations with the global context from the $[\text{CLS}]$ token. This concatenated vector is then passed through a feed-forward network (FFN) and a sigmoid function $\sigma$, as follows. By this means, the resulting score $s$, reflects a fine-grained semantic alignment, thereby holding the potential to address the semantic shortcut bias.
\[
s = \sigma(\text{FFN}(\text{Concat}(h_{\text{[CLS]}}, h''_{\text{[QSP]}}, h''_{\text{[PSP]}}))).
\]

\textbf{Hard Path Mining (HPM)}. Optimizing our cross-attentive architecture necessitates a training dataset. However, standard benchmarks typically lack ground-truth reasoning paths, requiring the self-construction of training instances. Consequently, previous studies often resort to using the single shortest path as a positive sample and random walks for negatives \cite{RD-P, luo2024reasoning, NEURIPS2024_efaf1c97}. Recognizing that these methods provide insufficiently challenging signals for fine-grained semantic learning, we employ the idea of the Hard Example Mining \cite{shrivastava2016training, smirnov2018hard, huang-etal-2024-selective-annotation} to facilitate our cross-attentive architecture's learning. Specifically, this involves constructing paths that are challenging for the retriever to accurately identify, thereby teaching it to discern beyond superficial textual similarity when using token-level interactions. As such, \ours~can be compelled to learn to distinguish between correct paths and semantically similar yet incorrect distractors. To this end, we propose a semantic-aware path sampling technique to explicitly construct these challenging examples prior to model training, encompassing both positive and negative paths. Refer to the pseudo-code in Algorithm \ref{alg:positive-path-sampling} and \ref{alg:negative-path-sampling} for details.

\begin{itemize}[leftmargin=*]
    \item \textit{Hard Positive Path Curation}. For each query $q$, multiple reasoning paths from the topic entity $e_t$ to the answer entity $e_a$ may exist in the KG $\mathcal{G}$. However, not all of them reflect the true reasoning logic. To ensure logical reasoning correctness and diversity, we first identify all shortest paths from $e_t$ to $e_a$. Since paths with low semantic similarity often lack logical relevance to the query, we then filter these to retain only those semantically aligned with the query's intent, using a SimCSE-based similarity score. Specifically, we compute cosine similarity $\text{sim}(q, p_i)$ for each reasoning path $p_i$ and the query $q$. Only paths with $\text{sim}(q, p_i) \geq \mu_q$ are selected as positive samples, where $\mu_q$ is the average similarity across all candidate paths for $q$:
    \[
    \mu_q = \frac{1}{|\mathcal{P}_q|} \sum_{p_i \in \mathcal{P}_q} \text{sim}(q, p_i).
    \]
    As such, this ensures our positive examples are challenging, factually correct, and logically relevant.

    \item \textit{Hard Negative Path Curation}. We generate negative paths by perturbing positive ones at each hop. Specifically, for a given positive path, if we are at step $i$ with relation $r_i$, we retain its prefix up to $i-1$ and replace $r_i$ with alternatives. These alternatives include: 1) \textit{Hard Negatives}. The Top-$K$ relations most semantically similar to $r_i$, which form plausible but logically invalid path extensions; 2) \textit{Normal Negatives}. $K$ relations randomly sampled from the entire relation set $\mathcal{R}$, introduced to enhance diversity. This dual strategy encourages the retriever to distinguish between semantically similar yet incorrect paths, thereby promoting deeper understanding beyond mere surface-level semantic similarity.
\end{itemize}
Leveraging these curated samples, which constitute our training instances, we formulate the following semantic discrimination loss to enhance the learning process of our token-level interaction. For a positive path $p_{\text{pos}}$ and a set of challenging negatives $\mathcal{N}_K$(comprising the $K$ highest-scoring incorrect paths selected from the generated hard and normal negatives), the objective is defined as:
\[
\mathcal{L}_{\text{SHC}} = -\log \frac{\exp(s_{\text{pos}})}{\exp(s_{\text{pos}}) + \sum_{j \in \mathcal{N}_K} \exp(s_j)},
\]
where $s$ represents the relevance score output by our retriever.

\begin{algorithm}[H]
\small
\caption{Hard Positive Path Curation}
\label{alg:positive-path-sampling}
\textbf{Input}: Query $q$, topic entities $\mathcal{E}_t$, answer entities $\mathcal{E}_a$, KG $\mathcal{G}$, max hop $H$, similarity model $M$
\begin{algorithmic}[1]
\STATE Initialize $\mathcal{P}_+ \gets \emptyset$
\FOR{each $e_t \in \mathcal{E}_t$}
    \STATE $\mathcal{P} \gets$ \texttt{Shortest\_Paths}($\mathcal{G}, e_t, \mathcal{E}_a, H$)
    \FOR{each $p \in \mathcal{P}$}
        \STATE $s(q, p) \gets$ \texttt{Sim}($M, q, p$)
    \ENDFOR
    \STATE $\mu_q \leftarrow \texttt{Mean}(\{s(q, p) \mid p \in P\})$;  $\mathcal{P}' \gets \{ p \in \mathcal{P} \mid s(q, p) > \mu_q \}$
    \STATE $\mathcal{P}_+ \gets \mathcal{P}_+ \cup \mathcal{P}'$
\ENDFOR
\STATE \textbf{return} Positive path set $\mathcal{P}_+$
\end{algorithmic}
\end{algorithm}

\begin{algorithm}[H]
\small
\caption{Hard Negative Path Curation}
\label{alg:negative-path-sampling}
\textbf{Input}: Positive paths $\mathcal{P}_+$, KG $\mathcal{G}$, similarity model $M$, \#negatives $K$
\vspace{-3mm}
\begin{algorithmic}[1]
\STATE Initialize $\mathcal{P}_- \gets \emptyset$
\FOR{each $p \in \mathcal{P}_+$}
    \FOR{each hop $(e, r^+) \in p$}
        \STATE $\mathcal{R} \gets$ \texttt{Relations}($\mathcal{G}, e$) $- \{r^+\}$
        \STATE $\mathcal{R}_{\text{hard}} \gets$ \texttt{TopK}($\mathcal{R},$ \texttt{Sim}($M, r^+, \mathcal{R}$), $K$)
        \STATE $\mathcal{R}_{\text{normal}} \gets$ \texttt{RandomK}($\mathcal{R}, K$)
        \STATE $\mathcal{P}_- \gets \mathcal{P}_- \cup \{(p, h, r^+, \mathcal{R}_{\text{hard}}, \mathcal{R}_{\text{normal}})\}$
    \ENDFOR
\ENDFOR
\STATE \textbf{return} Negative path set $\mathcal{P}_-$
\end{algorithmic}
\end{algorithm}

\subsection{Path-weighted Contrastive Learning} \label{sec:LOSS}
This section aims to fully unlock the potentials of \ours~via our customized contrastive learning. Unlike previous studies that treat all samples equally \cite{RD-P, NEURIPS2024_efaf1c97, hu-etal-2025-grag}, \ours~employs a customized graph context to weight each training path, thereby emphasizing rare, long-tail paths.

\textbf{Graph-Contextual Path Weighting}.
Prior to model training, \ours~explicitly weights each sample's contribution to the loss $\mathcal{L}_{\text{SHC}}$. This weighting method is designed to consider not only the path's individual rarity but also balances the overall influence between the positive and negative sample sets. Specifically, the raw weight $w_i$ for a path $p_i$ is determined by its inverse occurrence count ($1/n_i$), which is then scaled by the total number of paths in its respective class ($N_{\text{pos}}$ or $N_{\text{neg}}$) generated by semantic-aware path sampling within the training set. This class-wise scaling normalizes rarity scores and prevents one class from disproportionately dominating the loss. For training stability, these raw weights are subsequently scaled to a fixed range $[w^-, w^+]$, resulting in the final weight $\tilde{w}_i$. Incorporating these adjusted weights into the loss compels the model to allocate greater attention to rare positive paths, thereby enhancing robustness across the entire path distribution.

To tackle both biases simultaneously, we integrate this graph-contextual path weighting into our semantic-hard contrastive learning objective $\mathcal{L}_{\text{SHC}}$. The final loss function for \ours~is thus defined as $\mathcal{L}_{\text{\ours}}$, detailed as follows:
\[
\mathcal{L}_{\text{\ours}} = -\log \frac{\tilde{w}_{\text{pos}} \cdot \exp(s_{\text{pos}})}{\tilde{w}_{\text{pos}} \cdot \exp(s_{\text{pos}}) + \sum_{j \in \mathcal{N}_K} \tilde{w}_j \cdot \exp(s_j)}.
\]
Importantly, this unified formulation achieves two goals at once: the contrastive structure with hard negatives forces the model to learn deep semantics, while the weights $\tilde{w}$ ensure that this learning process is applied effectively across the entire data distribution, including the long tail. 

\textbf{Inference of \ours}. To further boost retrieval accuracy and efficiency, we follow previous works \cite{RD-P, sun2024thinkongraph} and employ a beam search inference process, where the trained \ours~serves as a scoring engine to guide a beam search over the KG. In particular, the process starts from the topic entity $e_t$ of the input query. At each step, we expand the current set of candidate paths by one more hop $r$ and use \ours~to score all new paths. Only the top-$K$ scoring paths are kept in the beam for the next expansion step. We introduce a special relation $r_{\text{[EOP]}}$ (End of Path) to determine path termination. The beam search terminates when expanding with $r_{\text{[EOP]}}$ yields the highest score. Finally, we pass the top-$K$ highest-scoring paths to the LLM, which selects the most relevant path to generate the answer. Pseudo-code of \ours~is in Algorithm \ref{alg:retrieval-generation}.

\begin{algorithm}[t]
\small
\caption{Inference Pipeline}
\label{alg:retrieval-generation}
\textbf{Input}: Query $q$, topic entities $\mathcal{E}_t$, knowledge graph $\mathcal{G}$, retriever model $M_s$, max hop $H$, path number $K$, beam width $B$, LLM $p_\phi$
\vspace{-3mm}
\begin{algorithmic}[1]
\STATE Initialize $\mathcal{C} \gets \emptyset$ \hfill // candidate paths
\FOR{each $e_t \in \mathcal{E}_t$}
    \STATE $\mathcal{Q} \gets$ \texttt{InitQueue}($e_t, [e_t], 1$)
    \WHILE{$\mathcal{Q}$ not empty}
        \STATE Pop $(e, p, s)$ from $\mathcal{Q}$  \hfill // entity, path, scores
        \IF{\texttt{HopCount}$(p) \geq H$}
            \STATE $\mathcal{C} \gets \mathcal{C} \cup \{(p, s)\}$; \textbf{continue}
        \ENDIF
        \STATE $\mathcal{R} \gets$ \texttt{Relations}($\mathcal{G}, e$) $ \cup \{\texttt{[EOP]}\}$
        \STATE $\mathcal{R}' \gets$ \texttt{Select}($M_s, q, \mathcal{R}$, $B$)
        \FOR{each $r \in \mathcal{R}'$}
            \IF{$r = \texttt{[EOP]}$}
                \STATE $\mathcal{C} \gets \mathcal{C} \cup \{(p \| r, s)\}$
            \ELSE
                \STATE $\mathcal{T} \gets$ \texttt{Tail}($\mathcal{G}, h, r$); $\mathcal{Q} \gets \mathcal{Q} \cup \{(t, p \| r, s) \mid t \in \mathcal{T}\}$
            \ENDIF
        \ENDFOR
    \ENDWHILE
\ENDFOR
\STATE $\mathcal{P} \gets$ \texttt{FinishedPaths}($\mathcal{C}$)
\STATE $p^* \gets$ \texttt{TopK\_Select}($\mathcal{P}, K$)
\STATE $a^* \gets$ \texttt{LLM\_Generate}($p_\phi, q, p^*$)
\STATE \textbf{return} Generated answer $a^*$
\end{algorithmic}
\end{algorithm}


\begin{table}[t]
\centering
\caption{Statistics, distribution of path depths, and hops.}
\label{tab:dataset-statistics}
\resizebox{0.45\textwidth}{!}{%
\begin{tabular}{lccccccc}
\toprule
\textbf{Dataset} & \textbf{\#Train} & \textbf{\#Dev} & \textbf{\#Test} & \textbf{Max Hop} & \textbf{1-hop} & \textbf{2-hop} & \textbf{$\geq$3-hop} \\
\midrule
WebQSP  & 2,543  & 283 & 1,628  & 2 & 65.49\% & 34.51\% & -  \\
CWQ     & 26,257 & 1,382 & 3,531  & 4  & 40.91\% & 38.34\% & 20.75\% \\
GrailQA & 3,800 & 200 & 1,000 & 4 & 68.93\% & 25.82\% & 5.25\%  \\
\bottomrule
\end{tabular}
}
\vspace{-3mm}
\end{table}
\section{Experiments}
Our evaluation first establishes its overall effectiveness by comparing it against strong baselines on benchmarks (See Section \ref{main}), which are derived from large-scale, Web-sourced knowledge graphs that simulate real-world Web interactions. Next, We conduct ablation studies with targeted analyses to isolate the individual contributions of \ours~and to confirm that \ours~successfully mitigates the identified the two biases (cf. Section \ref{ablation}).

\begin{table*}[htbp]
  \small
  \renewcommand{\arraystretch}{0.9}
  \centering 
  \caption{Performance comparison with different baselines on the benchmark datasets. \ours~consistently outperforms all baselines across various benchmarks in retrieval performance and often demonstrates superior retrieval efficiency. }
  \label{tab:performance_comparison}
  \begin{tabular}{p{3cm}|p{2.7cm}|*{9}{c}}
    \toprule
    \multirow{2}{*}{\textbf{Setting}} & \multirow{2}{*}{\textbf{Model}} & \multicolumn{3}{c}{\textbf{WebQSP}} & \multicolumn{3}{c}{\textbf{CWQ}} & \multicolumn{3}{c}{\textbf{GrailQA}} \\
    \cmidrule(lr){3-5} \cmidrule(lr){6-8} \cmidrule(lr){9-11}
    & & \textbf{Hits@1 (\%)} & \textbf{F1 (\%)} & \textbf{RT} & \textbf{Hits@1 (\%)} & \textbf{F1 (\%)} & \textbf{RT} & \textbf{Hits@1 (\%)} & \textbf{F1 (\%)} & \textbf{RT} \\
    \midrule
    \multirow{3}{*}{\textbf{LLM without KG}} & Qwen2.5-72B & 58.1 & 39.4 & 3.5 & 32.4 & 27.3 & 4.3 & 24.3 & 18.3 & 4.5 \\
    & Llama3.3-70B & 62.5 & 43.2 & 2.8 & 39.5 & 34.1 & 3.6 & 26.3 & 20.3 & 4.0 \\
    & DeepSeek-R1-70B & 75.2 & 59.0 & 16.3 & 55.3 & 50.3 & 17.2 & 41.9 & 34.2 & 17.8 \\
    \midrule
    \multirow{3}{*}{\textbf{LLM-centric Retriever}} & RoG & 87.1 & 71.6 & 1.6 & 83.3 & 64.8 & \underline{1.2} & 73.1 & 63.8 & \underline{2.3} \\
    & Graph CoT & 85.3 & \underline{73.5} & 123.3 & 66.1 & 61.4 & 513.6 & \underline{78.3} & \underline{73.1} & 168.4 \\
     & ToG & 67.6 & 61.3 & 41.7 & 52.1 & 43.1 & 90.4 & 69.8 & 60.4 & 62 \\
    \midrule
    \multirow{6}{*}{\textbf{Lightweight Retriever}} & DALK & 58.9 & 50.2 & 24.7 & 45.8 & 35.3 & 72.6 & 56.7 & 49.1 & 49.5 \\
    & G-Retriever & 63.1 & 69.3 & 8.9 & 42.1 & 40.8 & 6.7 & 48.5 & 49.8 & 7.7 \\
    & RD-P & \underline{88.1} & 73.1 & \textbf{0.8} & \underline{83.8} & \underline{66.2} & \textbf{1.1} & 71.3 & 67.9 & \textbf{1.1} \\
    & GRAG & 60.5 & 52.8 & 10.1 & 45.6 & 42.2 & 7.2 & 49.5 & 53.2 & 7.5 \\ 
    \cmidrule{2-11}
    & \ours~(\textit{Ours}) & \textbf{88.7} & \textbf{74.1} & \underline{0.9} & \textbf{86.3} & \textbf{69.1} & \underline{1.2} & \textbf{80.4} & \textbf{76.3} & \textbf{1.1} \\
    \bottomrule
  \end{tabular}
\end{table*}


\subsection{Experimental Setup}

\textbf{Datasets}. Aligning with the datasets detailed in Section \ref{psetup}, our experiments utilize three representative multi-hop benchmark datasets with large-scale KGs: \textit{WebQSP}~\cite{WebQSP}, \textit{CWQ}~\cite{CWQ}, and  \textit{GrailQA}~\cite{GrailQA}. 
These datasets, detailed in Table \ref{tab:dataset-statistics}, represent varying complexities and hop distributions (1-hop, 2-hop, and $\ge$3-hop) crucial for comprehensive evaluation.


\noindent\textbf{Evaluation Metrics.}
Following previous works \cite{RD-P, sun2024thinkongraph, NEURIPS2024_efaf1c97}, we employ \underline{Hits@1} and \underline{F1} as our primary evaluation metrics. In particular, Hits@1 verifies whether the top-ranked result matches any ground-truth answer entity. F1, on the other hand, evaluates the final generation by comparing the generated answer to all possible correct answers, thus balancing precision and recall. Additionally, we report \underline{RT (Retrieval Time)}, which measures the average time (in seconds) the retriever takes to process each query. This metric is crucial for assessing retrieval efficiency, particularly in real-world web applications where low-latency responses are critical.

\noindent\textbf{Baselines}. Following the setup detailed in Section \ref{psetup}, we consider all two groups of state-of-the-art GraphRAG baselines for comparison: (1) \underline{GraphRAG with LLM-centric retrievers}. These include \textit{RoG}~\cite{luo2024reasoning}, \textit{Graph CoT}~\cite{jin-etal-2024-graph}, and \textit{ToG}~\cite{sun2024thinkongraph}; (2) \uline{GraphRAG with Lightweight Retrievers}. These include \textit{DALK}~\cite{li-etal-2024-dalk}, \textit{G-Retriever}~\cite{NEURIPS2024_efaf1c97}, \textit{RD-P}~\cite{RD-P}, and \textit{GRAG}~\cite{hu-etal-2025-grag}. Moreover, we also consider the \uline{LLM without KG}, which assesses the performance of large-scale LLMs that directly answer queries without KG augmentation, including \textit{Qwen2.5-72B}~\cite{qwen2025qwen25technicalreport}, \textit{Llama3.3-70B}~\cite{meta2024llama33}, and \textit{DeepSeek-R1-70B}~\cite{guo2025deepseek}. 
{For a fair and reproducible comparison, all baseline models were implemented based on their officially released source code on github and followed the instructions provided by the original authors.}

\noindent\textbf{Implementation Details.}
All experiments are conducted on a machine equipped with an Intel(R) Xeon(R) Gold 6348 CPU, 1TB of RAM, and an NVIDIA A100 GPU. For baseline methods, we utilize their officially released open-source GitHub implementations.
Our proposed \ours, for the main experiments, is implemented using SimCSE backbone \cite{gao2021simcse} with a single-tower cross-attention architecture, trained using our custom loss function $\mathcal{L}_{\text{\ours}}$. {To handle the scale of Freebase, we extract local subgraphs for topic entities within dataset-specific hops. We adopt a unified sampling size of $K{=}15$ for hard path mining and set the graph-contextual loss weights to $w^+{=}3.0$ and $w^-{=}0.5$. The model is trained for 50 epochs on WebQSP/GrailQA (learning rate $3\text{e-}5$) and 100 epochs on CWQ ($1\text{e-}5$), utilizing a 90:10 and 95:5 train/validation split, respectively.}  
It is important to note that for all GraphRAG methods (including \ours), we apply beam search with a width of $K=3$ and select the top-1 path for answer generation. For fair and consistent comparisons across all evaluated methods, LLaMA3.3-70B serves as the answer generator, following common practice \cite{zhang2025survey}. 


\subsection{Main Results}
\label{main}
Table~\ref{tab:performance_comparison} summarizes our main results, demonstrating the superiority of \ours. Our key observations are as follows.

\noindent\textbf{\ours~consistently outperforms all baselines across various benchmarks in terms of the retrieval performance}. Table~\ref{tab:performance_comparison} clearly demonstrates the necessity of the GraphRAG paradigm for complex reasoning tasks, with LLM without KG baselines yielding the lowest scores. While LLM-centric retrievers such as RoG often establish upper performance limits (e.g., 87.1\% Hits@1 on WebQSP), they incur significant computational costs. Conversely, most Lightweight Retrievers prioritize efficiency but typically exhibit lower performance. This is precisely where \ours's contribution is critical. Importantly, \ours~consistently outperforms all baselines, achieving average retrieval performance gains (i.e., Hits@1) of 1.8\% and LLM QA performance improvements (i.e., F1) of 2.2\% across all evaluated datasets. For instance, on the complex CWQ dataset, \ours's state-of-the-art 86.3\% Hits@1 effectively provides the better context for the LLM, enabling it to achieve a leading 69.1\% F1 score. This performance gain is particularly significant within GraphRAG research, which has long struggled with a trade-off between performance and efficiency. Therefore, \ours~not only leads the lightweight category but also surpasses LLM-centric models (e.g., outperforming Graph CoT by 2.1 points on GrailQA). 

\noindent\textbf{\ours~often enjoys higher retrieval efficiency}. As evidenced in Table \ref{tab:performance_comparison}, \ours~achieves highly efficient retrieval times: ~1.0 seconds per query. This performance significantly surpasses LLM-centric retrievers, which incur substantially higher retrieval latencies (e.g., 513.6 seconds for Graph CoT on CWQ). Furthermore, \ours~outperforms most Lightweight Retrievers—including DALK, G-Retriever, and GRAG—in terms of retrieval efficiency, demonstrating superior speed while retaining competitive accuracy. This efficiency is largely attributable to direct path-scoring architecture, which avoids the bottlenecks present in other methods. For instance, DALK is slowed by its reliance on an LLM for costly post-retrieval filtering, while G-Retriever and GRAG require the exhaustive, unpruned encoding of all local graph components to construct a subgraph. Although \ours~sometime is marginally slower than RD-P, this slight trade-off is acceptable given its substantial improvements in both Hits@1 and F1 scores. To sum up, \ours~successfully bridges the performance-efficiency gap, offering heavy-weight system (i.e., LLM-based methods) accuracy with lightweight system efficiency, making it suitable for large-scale web applications where both speed and accuracy are paramount.

\subsection{In-depth Analysis \& Ablation Studies}
\label{ablation}
To further validate the effectiveness and characteristics of \ours, we conducted in-depth studies. These studies aim to reveal how \ours~succeeds by easing the two identified biases and to demonstrate its generalization across different PLM backbones. 

\noindent\textbf{Analysis Setup}. 
Aligning with the setup in Section \ref{psetup}, we evaluate results using four metrics. 1) \uline{Error}, defined as $1-Hits@1$, measures the overall retrieval failure rate. 2) \uline{Shortcut} represents the error rate primarily attributed to the semantic shortcut bias. 3) \uline{Long-Tail} quantifies the error rate stemming from the long-tail path bias. 4) Finally, \uline{Union} indicates the combined percentage of retrieval errors attributed to the presence of either of these two biases.

\begin{table}[h]
\centering

\setlength{\tabcolsep}{1pt}
\caption{Ablation analysis (\%) on \ours's token-level interaction learning to mitigate Semantic Shortcut Bias. Error percentages are shown outside parentheses, with the relative contribution to total errors inside.}
\label{tab:ablation-framework-merged}
\resizebox{0.44\textwidth}{!}{%
\begin{tabular}{l|l|cccc}
\toprule
\textbf{Backbone} & \textbf{Retriever Type} & \textbf{Error} & \textbf{Shortcut} & \textbf{Long-tail} & \textbf{Union} \\
\midrule
\multicolumn{6}{c}{\textsc{WebQSP}} \\
\midrule
\multirow{3}{*}{SimCSE} & Similarity-based & 14.8 & 6.8 (45.8) & 7.6 (51.6) & 10.4 (70.1) \\
& Cross-Attentive & \uline{12.6} & \uline{4.6 (36.7)} & \uline{6.7 (53.3)} & \uline{8.4 (66.7)} \\
& \ \ w/ HPM & \textbf{12.2} & \textbf{3.8 (31.3)} & \textbf{6.5 (49.1)} & \textbf{7.9 (66.2)} \\
\midrule
\multirow{3}{*}{TweetNLP} & Similarity-based & 16.4 & 4.9 (29.9) & 7.0 (42.7) & 9.5 (58.2) \\
& Cross-Attentive & \uline{12.4} & \uline{4.5 (36.2)} & \uline{6.3 (50.8)} & \uline{7.6 (63.3)} \\
& \ \ w/ HPM & \textbf{11.7} & \textbf{3.1 (26.7)} & \textbf{6.1 (50.1)} & \textbf{7.1 (60.5)} \\
\midrule
\multirow{3}{*}{BERT} & Similarity-based & 22.8 & 7.9 (34.6) & 11.4 (50.0) & 13.5 (59.3) \\
& Cross-Attentive & \uline{11.4} & \uline{4.3 (37.7)} & \uline{4.8 (42.1)} & \uline{7.0 (61.4)} \\
& \ \ w/ HPM & \textbf{10.2} & \textbf{2.9 (28.4)} & \textbf{4.3 (42.2)} & \textbf{6.5 (63.7)} \\
\midrule
\multicolumn{6}{c}{\textsc{CWQ}} \\
\midrule
\multirow{3}{*}{SimCSE} & Similarity-based & 24.0 & 15.3 (63.6) & 5.6 (23.5) & 17.3 (72.0) \\
& Cross-Attentive & \uline{14.8} & \uline{7.0 (47.2)} & \uline{3.8 (25.9)} & \uline{10.0 (67.6)} \\
& \ \ w/ HPM & \textbf{14.3} & \textbf{5.9 (41.0)} & \textbf{3.7 (25.8)} & \textbf{9.3 (64.9)} \\
\midrule
\multirow{3}{*}{TweetNLP} & Similarity-based & 22.1 & 11.4 (51.6) & 6.0 (27.3) & 15.1 (68.1) \\
& Cross-Attentive & \uline{16.1} & \uline{8.1 (50.3)} & \uline{5.5 (34.2)} & \uline{10.9 (67.7)} \\
& \ \ w/ HPM & \textbf{15.4} & \textbf{5.8 (37.7)} & \textbf{5.1 (33.1)} & \textbf{10.1 (65.9)} \\
\midrule
\multirow{3}{*}{BERT} & Similarity-based & 30.8 & 15.8 (51.3) & 7.4 (24.0) & 18.8 (61.2) \\
& Cross-Attentive & \uline{16.7} & \uline{8.9 (53.3)} & \uline{6.9 (41.3)} & \uline{11.0 (67.7)} \\
& \ \ w/ HPM & \textbf{15.8} & \textbf{6.9 (43.7)} & \textbf{6.4 (40.5)} & \textbf{10.5 (66.3)} \\
\bottomrule
\end{tabular}
}
\vspace{-3mm}
\end{table}


\subsubsection{Analysis on Token-Level Interaction Learning}
\label{ana1}
The token-level interaction learning in \ours~comprises two components: a cross-attentive architecture and hard path mining. This section assess if these components help mitigate the Semantic Shortcut Bias. More critically, we evaluate the generalization capabilities of our token-level interaction learning across different PLM backbones. For this analysis, we consider three distinct PLM backbones: \textit{SimCSE} \cite{gao2021simcse} (the backbone of RD-P \cite{RD-P}), \textit{TweetNLP} \cite{camacho-collados-etal-2022-tweetnlp} (a RoBERTa-based model for text classification), and \textit{BERT} \cite{DBLP:journals/corr/abs-1810-04805}. In this regard, for each backbone, we consider the following baselines and \ours~ablations. Note that to isolate the effects of our proposed loss $\mathcal{L}_{\text{\ours}}$, all models in this study are trained with a standard contrastive loss.
\begin{itemize}[leftmargin=*]
    \item \uline{Similarity-based}. Represents a standard similarity-based architecture, which exemplifies the dual-encoder paradigm commonly used by state-of-the-art methods \cite{RD-P}. Consistent with prior work \cite{RD-P, luo2024reasoning, NEURIPS2024_efaf1c97}, we construct the training dataset using single shortest paths as positive samples and random walks as negatives. 
    \item \uline{Cross-attentive}. This variant substitutes the similarity-based architecture with our proposed cross-attentive architecture, utilizing the identical training dataset as the \textit{Similarity-based}.
    \item \uline{w/ HPM}. Further refines the \textit{Cross-attentive} by integrating hard path mining (HPM), training with our specially curated hard path examples. This configuration represents the full token-level interaction learning approach of \ours.
\end{itemize}


\noindent\textbf{Our token-level interaction learning mitigates Semantic Shortcut Bias, contributing to \ours's overall success}. Table~\ref{tab:ablation-framework-merged} demonstrates the effectiveness of our token-level interaction learning and its generalization capabilities across different PLM backbones.
Regarding the \uline{cross-attentive architecture}, it consistently reduces errors by an average of 5.9\% on WebQSP and 9.8\% on CWQ across various backbones compared to standard similarity-based architectures. Without explicit query-path interaction, standard similarity-based architectures confirm their vulnerability to failure modes. For instance, shortcut errors constitute a staggering 45.8\% (6.8\%/14.8\%) of total failures on WebQSP. This problem is exacerbated on the more complex CWQ dataset, where shortcut errors account for a majority 63.6\% (15.3\%/24.0\%) of all retrieval failures. This trend underscores that as reasoning complexity increases, reliance on superficial similarity leads to degraded performance. In contrast, our cross-attentive architecture significantly mitigates this specific bias, leading to a reduction in shortcut errors by 1.8 points on WebQSP and a remarkable 8.3 points on CWQ. For better understanding, a case study, presented in Section ~\ref{case_study}, visualizes our model's token-level attention distribution. 
Concurrently, Table \ref{tab:ablation-framework-merged} indicates that our \uline{hard path mining} further yields an average error reduction of 0.8\% on WebQSP and 0.7\% on CWQ across all backbones, compared to baselines. Notably, we also observe that similarity-based retrievers show strong dependence on sentence-level semantic encoders. When using SimCSE, the similarity-based retriever attains relatively moderate errors (14.8\% on WebQSP, 24.0\% on CWQ). However, when replaced with BERT, which is not optimized for sentence-level representation learning, the error sharply rises to 22.8\% and 30.8\%, respectively. This reveals that the dual-encoder paradigm heavily relies on backbone semantics to align queries and paths. In contrast, our cross-attentive and hard path mining variants maintain stable errors ($\sim$10–16\%) across all backbones, demonstrating that token-level interaction learning reduces such dependence and can achieve backbone-agnostic robustness.


\begin{table}[h]
\centering
\caption{Ablation analysis (\%) on \ours's path-weighted contrastive learning. Error percentages are shown outside parentheses, with the relative contribution to total errors inside.}
\label{tab:ablation-loss-merged}
\resizebox{0.45\textwidth}{!}{%
\begin{tabular}{l|cccc}
\toprule
\textbf{Objective} & \textbf{Error} & \textbf{Shortcut} & \textbf{Long-tail} & \textbf{Union} \\
\midrule
\multicolumn{5}{c}{\textsc{WebQSP}} \\
\midrule
Contrastive Loss & 12.6 & 4.6 (36.7) & 6.7 (53.3) & 8.4 (66.7) \\
\ \ \ \ w/ Weight & \underline{11.9} & \underline{4.0 (33.3)} & \textbf{5.8 (49.1)} & \underline{7.9 (66.2)} \\
\ours & \textbf{11.3} & \textbf{3.7 (32.7)} & \underline{5.8 (51.1)} & \textbf{7.4 (65.1)} \\
\midrule
\multicolumn{5}{c}{\textsc{CWQ}} \\
\midrule
Contrastive Loss & 14.8 & 7.0 (47.2) & 3.8 (25.9) & 10.0 (67.6) \\
\ \ \ \ w/ Weight & \underline{14.0} & \underline{6.2 (44.3)} & \textbf{3.2 (23.0)} & \underline{8.9 (63.7)} \\
\ours & \textbf{13.7} & \textbf{5.8 (42.4)} & \underline{3.3 (24.1)} & \textbf{8.7 (63.4)} \\
\bottomrule
\end{tabular}
}
\label{tab:ablation-loss-merged}
\end{table}

\subsubsection{Analysis on Path-weighted Contrastive Learning}

To verify that our path-weighted contrastive learning effectively mitigates the Long-Tail Path Bias, we dissect the significance of the path weight by comparing three training objectives. In this regard, we consider the following baselines and \ours~ablations. Note that all methods are applied to our cross-attentive retriever to isolate the effects of the loss function.
\begin{itemize}[leftmargin=*]
    \item \uline{Contrastive Loss}. This variant is the \textit{Cross-attentive} in Section \ref{ana1}, employing a standard contrastive loss function with randomly sampled negatives.
    \item \uline{w/ Weight}. Refines the \textit{Contrastive Loss} by integrating our graph-contextual path weighting.
    \item \uline{\ours}. Represents the full proposed method, further enhancing the \textit{w/ Weight} configuration with hard path mining.
\end{itemize}

\noindent\textbf{Our path-weighted contrastive learning mitigates both Long-tail Path Bias}.
As shown in Table~\ref{tab:ablation-loss-merged}, By amplifying the learning signal for infrequent paths, \ours's weighting mechanism forces the model to pay more attention to rare reasoning patterns. This shows success, lowering the long-tail errors from 6.7\% to 5.8\% on WebQSP, and from 3.8\% to 3.2\% on CWQ. This directly confirms its effective utility in improving robustness on skewed data distributions. Notably, our full version of \ours~further reduces shortcut errors while maintaining strong gains on long-tail paths, keeping these errors as low as 5.8\% and 3.3\% on WebQSP and CWQ, respectively. This confirms a synergistic effect, validating that the components of our proposed loss work in concert to provide a comprehensive solution for both identified biases.

\subsubsection{Analysis on Robustness to Different Answer Generators}
The retriever's output is subsequently fed into a downstream LLM to generate the final answer to the user query. This section further validates \ours's robustness across different answer generators, thereby confirming the high quality of its retrieval results. For this validation, we consider three representative LLMs: \textit{LLaMA3.3-70B}, \textit{Qwen2.5-72B}, and \textit{DeepSeek-R1-70B}. Across all experimental settings, unless otherwise specified, only the Top-1 reasoning path from \ours~is utilized by the LLM.

\begin{table}[h]
\centering

\setlength{\tabcolsep}{9pt}
\renewcommand{\arraystretch}{1.1}
\setlength{\abovecaptionskip}{0pt}
\setlength{\belowcaptionskip}{0pt}
\caption{F1 (\%) performance of \ours~with different LLM answer generators. Performance improvements over their respective counterparts in Table \ref{tab:performance_comparison} are highlighted in brackets.}
\resizebox{0.49\textwidth}{!}{%
\begin{tabular}{l|ccc}
\toprule
\textbf{Model} & \textbf{WebQSP} & \textbf{CWQ} & \textbf{GrailQA} \\
\midrule
\textbf{LLaMA3.3-70B} & \underline{75.3 (+35.9)} & \underline{69.7 (+42.4)} & \underline{77.2 (+58.9)} \\
\textbf{Qwen2.5-72B} & 72.3 (+29.1) & 68.8 (+34.7) & 76.7 (+56.4)\\
\textbf{DeepSeek-R1-70B} & \textbf{86.3 (+27.3)} & \textbf{70.0 (+19.7)} & \textbf{78.2 (+44.0)} \\
\bottomrule
\end{tabular}
}
\label{tab:llm-variants}
\end{table}

\noindent\textbf{\ours~consistently provides high-quality paths that robustly enhance various LLMs, achieving top-tier performance}. Table~\ref{tab:llm-variants} highlights the universal effectiveness of \ours~as a robust enhancement for various LLMs, evidenced by the consistent F1 score uplift observed across all tested LLMs. For instance, when paired with \ours, DeepSeek-R1-70B consistently yields the best results, pushing the F1 score to a remarkable 86.3\% on WebQSP. This robust performance confirms that \ours~produces high-quality paths, establishing \ours~as a powerful and generalizable tool for diverse downstream LLMs. Furthermore, according to the Table \ref{tab:topk-llm}, the answer generation performance of LLMs progressively improves as the number of Top-K retrieved paths increases. While $K=1$ achieves strong performance, the Hits@1 score reaches an impressive 95.2\% on WebQSP at $K=7$. This confirms \ours's high effectiveness in retrieving relevant paths and placing the correct answer among the top candidates. The notable gap between $K=1$ and higher $K$ values indicates that, despite its strong performance, the retriever can still rank plausible but incorrect paths above the correct ones. This underscores the persistent challenge posed by the two biases and underscores the need for future work to refine the retriever's discrimination against subtle distractors. 

\begin{table}[h]
\centering

\renewcommand{\arraystretch}{1.1}
\setlength{\tabcolsep}{12pt}
\setlength{\abovecaptionskip}{0pt}
\setlength{\belowcaptionskip}{0pt}
\caption{Hits@1 (\%) of \ours~under different top-$K$ settings.}
\resizebox{0.4\textwidth}{!}{%
\begin{tabular}{l|cccc}
\toprule
\textbf{Dataset} & $K$=1 & $K$=3 & $K$=5 & $K$=7 \\
\midrule
\textbf{WebQSP} & 88.7 & 93.9 & 95.0 & \textbf{95.2} \\
\textbf{CWQ} & 86.4 & 87.3 & 89.0 & \textbf{89.4} \\
\textbf{GrailQA} & 80.1 & 86.5 & \textbf{89.8} & \textbf{89.8} \\
\bottomrule
\end{tabular}
}
\label{tab:topk-llm}
\end{table}


\subsection{Case Study Analysis}
\label{case_study}
\begin{figure}[t]
  \centering
  \includegraphics[width=0.48\textwidth]{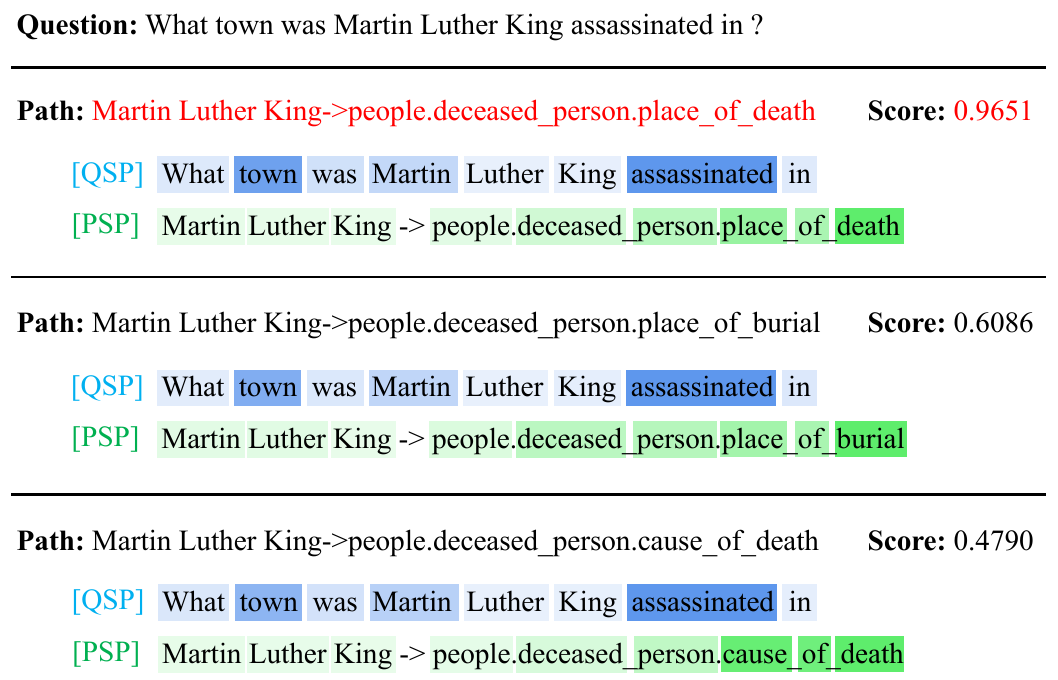}
  \setlength{\abovecaptionskip}{0pt}
\setlength{\belowcaptionskip}{0pt}
  \caption{Cross-Attention Visualization. 
  Darker color indicates stronger attention. \ours~assigns the highest score to the logically correct path \texttt{place\_of\_death} by aligning logically relevant tokens such as \texttt{town} and \texttt{assassinated}.}
  \label{fig:cross-attention-case}
  \vspace{-3mm}
\end{figure}

To further illustrate the inner workings of our Cross-Attentive Architecture and its effectiveness in mitigating \textit{Semantic Shortcut Bias}, we conduct a case study by visualizing the attention distribution in the cross-attention module. Specifically, we visualize the attention weights between the query special token \texttt{[QSP]} and the path special token \texttt{[PSP]}, shown in Figure~\ref{fig:cross-attention-case}. Darker colors indicate stronger attention. It visualizes the attention patterns between \texttt{[QSP]} and \texttt{[PSP]} for three candidate paths. In these paths, the baseline retriever incorrectly ranks the \texttt{place\_of\_burial} path higher than the correct \texttt{place\_of\_death} path due to high textual overlap. In our model, attention from \texttt{[PSP]} captures the key inferential clue in each candidate: the final relation token (e.g., \texttt{place\_of\_death}, \texttt{place\_of\_burial}, \texttt{cause\_of\_death}), which directly determines the logical correctness of the path. This reveals how the model focuses on key semantic alignments, distinguishing logically correct paths from distractors with high textual similarity.

To examine the behavior of various retrieval methods, we fed the query from Figure \ref{fig:cross-attention-case} (`\textit{What town was Martin Luther King assassinated in?}') into several baselines. Specifically, 
\uline{RD-P}, relying on similarity-based scoring, suffers from Semantic Shortcut Bias, incorrectly favoring the path \texttt{Martin Luther King} $\rightarrow$ \texttt{people.deceased\_\allowbreak person.place\_\allowbreak of\_\allowbreak burial} $\rightarrow$ \texttt{Atlanta} due to high surface-level semantic overlap, while \uline{Graph CoT} struggles with the excessive number of relations connected to \texttt{Martin Luther King}, leading it to hallucinate a non-existent path. Similarly, \uline{GRAG} also erroneously selects the relation \texttt{people.deceased\_\allowbreak person.place\_\allowbreak of\_\allowbreak burial}. \uline{\ours} effectively mitigates this bias by leveraging token-level interaction learning to focus attention on ``town'' and ``assassinated'' in the query and critical path segments, thereby accurately identifying \texttt{Martin Luther King} $\rightarrow$ \texttt{people.deceased\_\allowbreak person.place\_\allowbreak of\_\allowbreak death} $\rightarrow$ \texttt{Memphis}, moving beyond superficial similarity toward deeper logical reasoning, leading to correct retrieval.

\section{Conclusion}
In this work, we identified two critical biases in Lightweight Retrievers for GraphRAG: the Semantic Shortcut Bias and the Long-Tail Path Bias. These biases, we posit, stem from existing methods' oversight of the sparse semantic challenges inherent in graph structures. To address these limitations, we proposed \ours, a novel method whose experimental validation demonstrated superior performance, surpassing both existing lightweight and LLM-centric retrieval approaches. While our analysis confirms that \ours~effectively mitigates these two biases, thereby contributing to its success, this work primarily serves as a foundational step. Future research can build upon our insights by extensively exploring advanced graph-aware mechanisms and representation learning techniques that fully leverage sparse graph semantics, ultimately fostering more robust and efficient retrieval within complex knowledge graphs. 

\clearpage

\bibliographystyle{ACM-Reference-Format}
\bibliography{www2026}


\appendix

\end{document}